\documentclass{article}
\usepackage[utf8]{inputenc}
\usepackage{bbm}
\usepackage{empheq}
\usepackage{amsmath}
\usepackage{graphicx}
\usepackage{tcolorbox}
\usepackage{comment}
\usepackage{biblatex}
\usepackage{authblk}

\newtheorem{assumption}{Assumption}[section]

\title{Risk scoring calculation for the current NHSx contact tracing app}

\author[1, *]{Mark Briers}
\author[1]{Marcos Charalambides}
\author[1, 2]{Chris Holmes}

\affil[1]{\small The Alan Turing Institute, London, U.K.}
\affil[2]{\small University of Oxford, Oxford, U.K.}
\affil[*]{\small Corresponding author: Mark Briers, mbriers@turing.ac.uk}

\date{\today}

\begin{document}

\maketitle

\section{Introduction}

We consider how the NHS COVID-19 application will initially calculate a risk score for an individual based on their recent contact with people who report that they have coronavirus symptoms.

The NHS COVID-19 app uses Bluetooth to estimate the distance over time between people who have downloaded and are running the app. If a person reports coronavirus symptoms, their recent history of interactions is uploaded to a database and the risk scoring algorithm is used to update the risk score for every app user they have come into contact with.

The NHSx technical report \cite{riskscoreinfo} gives a high-level overview of the risk score calculation suitable for a wide audience. The report \cite{fraser2020contactv2} presents the algorithmic elements of the risk score calculation and notification process.

In this note, we describe the technical aspects of the risk scoring algorithm and consider its statistical basis.

\section{Risk score calculation}

Assume that individual $i$ notifies the app that they are symptomatic. The time of the reported onset of their symptoms is denoted as $t_{i}^{s}$ and this is distinct from the time of notification, denoted here by $t_{i}^{r}$. The superscripts $s$ and $r$ refer to the onset of symptoms and reporting time respectively. Currently, $t_i^s$ is always marked to noon of the day of symptom onset. All times are measured in minutes, unless otherwise specified.

Assume that the app for individual $i$ has stored $N_{i}$ contact events (somewhat ambiguously referred to as ``pings'' in \cite{fraser2020contactv2}). We will denote the $m$-th contact event as $E_{m}^{i}$ and the absolute time of the start of the contact event will be denoted by $t_{i,m}$. We will refer to $i$ as the \emph{source}  and the individuals associated with recorded contact events as \emph{recipients}.

Each contact event $E_m^i$, has an associated risk of transmission which can be written as a product as follows:
\begin{equation}\label{eqn:indiv-risk-score}
r(E_m^i)=\alpha^{t_i^s}_i\times c_{i,m} \times D_{i,m} \times I_{i,m} \times \delta t_{i,m}
\end{equation}
where:
\begin{itemize}
    \item $\alpha_{i}^{t}$ is a weighting associated with attributes of the source individual $i$ at time $t$ (such as severity of symptoms, age, etc) - currently $\alpha_{i}^{t}\equiv 1$;
    \item $c_{i,m}$ is a risk context adjusting factor, for example, taking into account factors such as whether the contact is made indoors, etc;
    \item $D_{i,m}$ is a distance-related risk factor;
    \item $I_{i,m}$ is a infectiousness risk factor;
    \item $\delta t_{i,m}$ is a duration of the contact event.
\end{itemize}

The distance-related factor is given by
\begin{equation}
    D_{i,m} = \min\left(1,\frac{d_{min}^2}{d_{i,m}^2}\right)
\end{equation}
where $d_{i,m}$ denotes the distance between source and recipient for the contact event and $d_{min}$ is a parameter controlling the point where the distance-related factor is maximised - currently $d_{min}=1$. 

The distance $d_{i,m}$ is a function of the Bluetooth signal which may be estimated using the RSSI (Received Signal Strength Indicator) value; see, for example, Equation 1 in \cite{rodas2008bayesian}.

The infectiousness factor is given by
\begin{equation}\label{eqn:infectiousness-gaussian}
    I_{i,m} = \exp\left(-\frac{1}{2}\left(\frac{([t_{i,m}]_{days}-[t_i^s]_{days} - \mu_0)}{\sigma_0}\right)^2\right)
\end{equation}
where $[t_{i,m}]_{days}-[t_i^s]_{days}$ denotes the time difference in days between the start of the contact event and midday of the day of symptom onset of source $i$ and the parameters $\mu_0$ and $\sigma_0$ control the shape of the Gaussian - currently $\mu_0=-0.3$, $\sigma_0=2.75$ (see Section \ref{sec:infectiousness-factor} for a discussion on these parameter values).

See Appendix \ref{sec:visualisations-risk} for a visualisation of the risk score $r(E_m^i)$.
    
The total risk of transmission, $r_{i,j}$, from source $i$ to a recipient $j$ is then obtained by aggregating the risk of transmission of all relevant contact events:

\begin{equation}\label{eqn:risk-score}
r_{i,j} = \sum_{m\,:\,id(E_m^i)=j} \mathbbm{1}(t_{i}^{s}-\Delta t_{max}< t_{i,m}) r(E_m^i)
\end{equation}
where:
\begin{itemize}
    \item $id(E_m^i)$ is the recipient identifier of the $m$-th contact event for source $i$;
    \item $\Delta t_{max}$ corresponds to the maximum amount of time a contact event is stored before the onset of symptoms of the source - currently this is $7$ days, i.e. $\Delta t_{max}=10080$.
\end{itemize}

Note that, once a source $i$ notifies the app, they do not continue to upload contact events. This implicitly assumes that the source $i$ self-isolates after the time of notification $t_i^r$.

\begin{assumption}
The source $i$ is not involved in any contact events after time $t_i^r$.
\end{assumption}

Finally, the risk score, $r_j$, associated to individual $j$ is obtained by aggregating the risk of transmission from all source individuals:
\begin{equation}
    r_j = \sum_{i}r_{i,j}.
\end{equation}

\subsection{Infectiousness factor}\label{sec:infectiousness-factor}

The \emph{generation period} between a source infector and a recipient is defined as the interval between the source becoming infected and the target becoming infected. The \emph{incubation period} is defined as the interval between the time an individual becomes infected and the time the individual shows symptoms. 

The infectiousness factor adjusts the risk score so that contact events on the day the source develops symptoms have higher weight. It accounts for the non-uniformity in the distribution of the time between a source infector showing symptoms and a contact event which causes a recipient to become infected. In fact, \cite{ferretti2020quantifying} model the incubation and generation periods of the virus and we can use these to estimate this distribution. The distribution ends up being numerically close to the normal distribution $N(-0.3, 2.75^2)$. We perform this analysis below. The Gaussian infectiousness factor in Equation \ref{eqn:infectiousness-gaussian} is equal to the density of this normal distribution which is scaled so that the maximum value is $1$.

In \cite{ferretti2020quantifying}, the incubation period is modelled by a log-normal distribution (following \cite{lauer2020incubation}) and the generation period is modelled by a Weibull distribution.

When the source $i$ uploads their data, we do not observe their time of infection but only the onset time of their symptoms, $t_i^s$. In practice, this time has uncertainty associated with it, but it is assumed to be negligible for the calculation of the risk score.

\begin{assumption}
    The uncertainty in the reported $t_i^s$ is negligible.
\end{assumption}

The infectiousness factor corresponds to modelling the time between the source's onset of symptoms and the potential contact time when the recipient becomes infected. This is precisely the distribution of the difference between the generation period and the incubation period.

\begin{assumption}
    The incubation and generation periods are independent.
\end{assumption}

\begin{figure}
    \centering
    \includegraphics[width=1\textwidth]{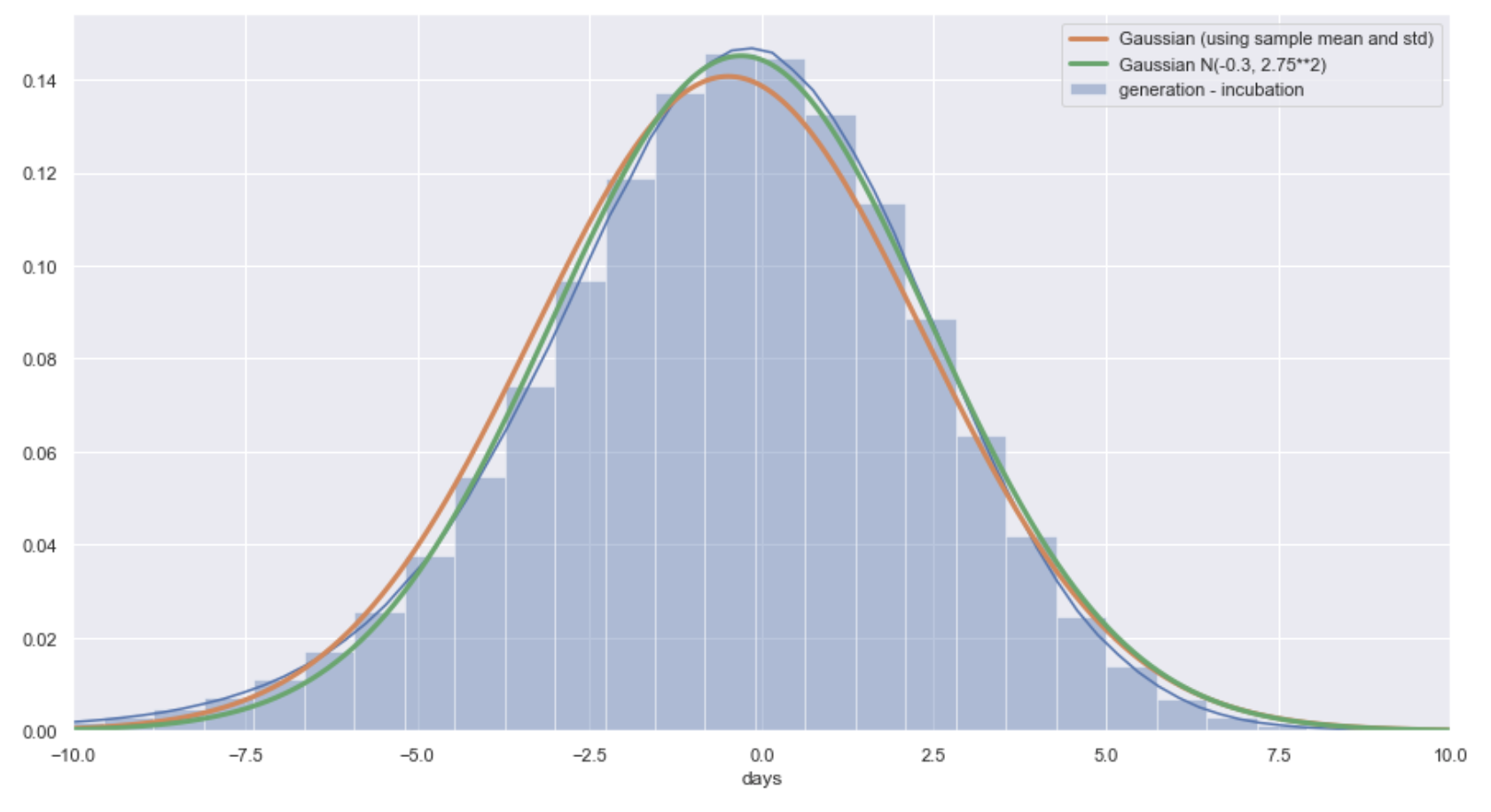}
    \caption{Infectiousness factor}
    \label{fig:infectiousness}
\end{figure}

In Figure \ref{fig:infectiousness}, we use the parameter estimates from \cite{ferretti2020quantifying} for the generation and incubation periods and generate samples for the difference. This is compared to the Gaussian $N(\cdot;\mu_0, \sigma_0^2)$.

The sample distribution is not symmetric and has a heavier left tail than the Gaussian. However, the Gaussian approximation is numerically close.

\section{Notification}

An individual $j$ is notified if their risk score is greater than or equal to a minimal risk score,
\[
r_{j} \geq r_{min}.
\]
A notified individual is advised to follow advice related to additional restrictions for a fixed period of $14$ days.

Currently, $r_{min}=1.83$ which is chosen to correspond to the PHE guidelines of a contact event of $15$ minutes at $2$ metres and marked to $3$ days from when the source develops symptoms.

\subsection{De-cascading}

If a source individual $i$ returns a negative test result, then their proximity risk at times before the test can be assumed to be equal to $0$ (assuming the tests have a low false negative rate), therefore a simple approach would be to adjust the source to receiver risk scores to 0. For each of the previously notified contacts, the de-cascading process should compute the revised risk scores for this individual, and if they fall below the threshold, notify them that they no longer need to follow advice related to additional restrictions. That is, one needs to ensure that the default de-cascading process does not release recipients whose risk score remains above $r_{min}$ after removing the risk component $r_{i,j}$.

\section{A probabilistic interpretation}\label{sec:prob-interpretation}

\subsection{The need for a probabilistic model}

While the risk score definition allows for a scalar valuation of infection risk to be computed, the fact that there is no underlying probabilistic model presents several challenges.

The meaning of the risk score values themselves lacks clarity and it becomes difficult to compare events directly. Moreover, there are a number of parameters in the model and there is no natural loss function which can be utilised in updating these parameters as data is received.

Another more practical problem is that once a recipient is notified, they follow advice for $14$ days regardless of whether the actual probability of them having been infected has significantly decreased during that period of time.

In this section, we give one possible interpretation which ties the risk score to the probability of infection. This allows us to present an approach which can address the aforementioned challenges.

\subsection{Risk score}

For an individual $j$, let $(E_n)_{n=1}^{N_{j}}$ be the sequence of contact events from symptomatic source individuals which have $j$ as recipient and are within the $\Delta t_{max}$ (i.e. $14$ day) cutoff. Write $I_E$ for the event that $j$ gets infected as a result of contact event $E$. Write 
\begin{equation}
    I_{j} = \bigcup_{n=1}^{N_{j}}I_{E_n}
\end{equation}
for the event that the recipient $j$ gets infected as a result of any of the contact events $(E_n)_n$.

We seek to relate the risk score $r_{j}$ to the probability that $j$ gets infected $I_{j}$ so that a higher risk score directly corresponds to a higher infection probability. The formulation should respect the fact that $r_{j}$ is the sum of the individual risk scores $(r(E_n))_n$ associated to the contact events.

We assume that the infection events $(I_{E_n})_n$ are independent and interpret the risk score $r(E)$ of the contact event $E$ as the negative of the logarithm of the conditional probability that $j$ does not get infected. 

\begin{assumption}
    The infection events $(I_{E_n})_{n=1}^{N_{j}}$ are (pairwise) independent.
\end{assumption}

More precisely, let $\nu\in (0, 1)$ be a parameter. Define $\rho(E)$ by
\begin{equation}\label{eqn:rho-event}
    \mathbbm{P}(\overline{I_E}) = \nu^{\rho(E)}.
\end{equation}
Here, $\overline{I_E}$ denotes the complement of the event $I_E$.

Define $\rho_{j}$ by
\begin{equation}
    \rho_{j}=\sum_{n=1}^{N_{j}}\rho(E_n).
\end{equation}
Then,
\begin{equation}
\begin{split}
    \mathbbm{P}(I_{j}) &= 1-\mathbbm{P}(\overline{I_{j}}) \\
    &= 1-\mathbbm{P}(\bigcap_n\overline{I_{E_n}}) \\
    &= 1-\prod_n\nu^{\rho(E_n)} \\
    &= 1 - \nu^{\rho_{j}}.
\end{split}
\end{equation}

We can view the risk score $r(E)$ (from equation \ref{eqn:indiv-risk-score}) as an estimator for $\rho(E)$ in equation \ref{eqn:rho-event} and then the risk score $r_{j}$ (from equation \ref{eqn:risk-score}) becomes an estimator for $\rho_{j}$ and has a clear probabilistic interpretation.

\subsection{Notification}

With this formulation, the notification process can be formulated in probabilistic terms.

Decide on a probability threshold $p_{min}$. An individual $j$ is notified if the probability that they have been infected given their contact events is equal to or above this threshold. That is, if
\begin{equation}
    \mathbbm{P}(I_{j})=1-\nu^{\rho_{j}} \geq p_{min}.
\end{equation}
In particular, the probability of a false positive is bounded above by $1-p_{min}$.

Consider, as above, the sequence of contact events $(E_n)_{n=1}^{N_j}$ for a recipient $j$. We will denote the time of the contact event $E$ by $t_E$. We will also denote by $S(t)$ the event that $j$ develops symptoms by time $t$ due to the contact events.

Recall that the generation period is the time between the source becoming infected and a recipient becoming infected and the incubation period is the time between an individual becoming infected and developing symptoms.

If the individual becomes infected as a result of contact event $E$, they will develop symptoms precisely after the generation period and the incubation period of the virus elapse from the time, $t_E$, that the contact event $E$ occurs. This means that the probability the recipient shows symptoms by time $t$ is equal to the probability that the sum of the generation period and the incubation period of the virus is at most the time elapsed since the contact event occurred, $t-t_E$.

Therefore, we can write
\begin{equation}
    \mathbbm{P}(S(t) | I_E)=G(t-t_E)
\end{equation}
where $G$ is the cumulative distribution function of the distribution of the sum of generation period and incubation period. 

Currently, a notified individual must follow additional advice for $14$ days. However, we may amend this algorithm so that a notified individual is subsequently informed that they no longer need to follow the additional restrictions when the probability $\mathbbm{P}(I_{j} | \overline{S(t)})$ that they are infected given that they have not experienced symptoms by time $t$ drops below a certain threshold. This probability can be expressed in terms of $\rho$, if we make the following assumption.

\begin{assumption}
    The events $\left(I_{E_n}\cap\overline{S(t)}\right)_{n=1}^{N_{j}}$ are pairwise independent.
\end{assumption}

Then, the probability that individual $j$ is infected without showing symptoms by time $t$ may be expressed as
\begin{equation}\label{eqn:rho-symptom}
\begin{split}
    \mathbbm{P}(I_{j} | \overline{S(t)}) 
    & = \frac{1 - \prod_{n}\left(1 - \left( 1 - G(t-t_{E_n})\right)\left(1 - \nu^{\rho(E_n)}\right)\right)}{1-\sum_n G(t-t_{E_n})(1-\nu^{\rho(E_n)})}.
\end{split}
\end{equation}
See Appendix \ref{sec:symptom-derivation} for a derivation.

The incubation period model in \cite{ferretti2020quantifying} (and \cite{lauer2020incubation}) is log-normal. As such, it assumes that there is $0$ probability of being asymptomatic.
With the models in \cite{ferretti2020quantifying} for incubation and generation periods, as $t\to\infty$ the cumulative distribution 
\[G(t-t_E)\to 1.\]
Therefore, the probability of being infected but never showing symptoms is zero, 
\[\mathbbm{P}(I_{j} | \overline{S(t)})\to 0.\]

\begin{figure}
    \centering
    \includegraphics[width=1\textwidth]{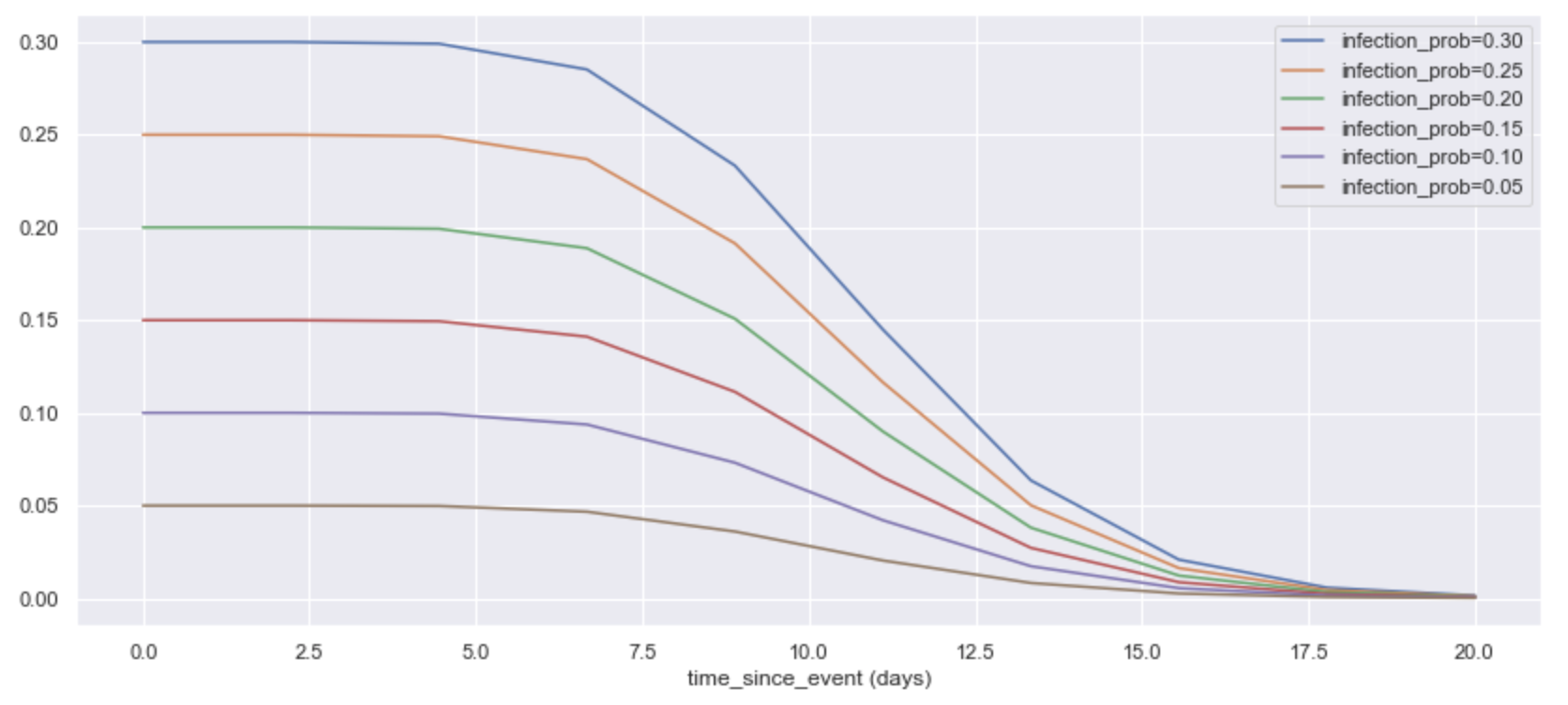}
    \caption{Probability recipient is infected without showing symptoms}
    \label{fig:prob-decay}
\end{figure}

In the case when there is just a single contact event $E$, the expression for the probability reduces to 
\begin{equation}
\begin{split}
    \mathbbm{P}(I_{j} | \overline{S(t)}) 
    & = \frac{\left( 1 - G(t-t_{E})\right)\left(1 - \nu^{\rho(E)}\right)}{1-G(t-t_{E})(1-\nu^{\rho(E)})}.
\end{split}
\end{equation}
Figure \ref{fig:prob-decay} plots the probability $\mathbbm{P}(I_{j} | \overline{S(t)})$ as a function of the time interval $t-t_E$ (\verb|time_from_event|) for a range of values of $\mathbbm{P}(I_{j})\equiv 1-\nu^{\rho(E)}$ (\verb|infection_prob|) using the distributions from \cite{ferretti2020quantifying}.

\subsection{Estimating parameters}

The probabilistic interpretation gives an approach to updating parameters. As an example, we consider how to estimate the parameter $\nu$.

Suppose we observe $M$ independent collections $\mathcal{D}=(C_{j_m})_{m=1}^M$ of contact events of recipients $j_m$. 

For each collection of contact events $C_{j_m}$, write $o_m$ for the observation of whether the individual $j_m$ becomes infected; so $o_m=1$ if $j_m$ gets infected and $o_m=0$ if $j_m$ does not get infected.

Assume a $Beta(1,1)$ prior for $\nu$, so $p(\nu)\equiv1$. By Bayes' rule,
\begin{equation}
\begin{split}
    p(\nu | \mathcal{D}) &\propto p(\nu)\prod_{m=1}^M\mathbbm{P}(o_m; C_{j_m} | \nu) \\
    &=\prod_{m=1}^M \nu^{\rho_{j_m}(1-o_m)}(1 - \nu^{\rho_{j_m}})^{o_m}.
\end{split}
\end{equation}

We can estimate $\rho_{j_m}$ by $r_{j_m}$ and apply MCMC, for example, to approximate the posterior distribution $p(\nu | \mathcal{D})$. Moreover, we can update this distribution as new data is received.

\section{Next steps}

There are several directions for further work relating to the risk score calculation as data is collected:

\begin{itemize}
    \item Improve estimation of the Bluetooth RSSI to distance mapping to increase the accuracy of the derived distance. This should enable the quantification of the uncertainty in the derived distance and how this uncertainty propagates through any calculations.
    \item Analyse the rate of false positives the application generates when notifying users at different risk score levels.
    \item Account for the false negative rate of the COVID-19 test (as in a user has tested negative despite still having the disease).
    \item Account for the uncertainty in the parameter estimates of the models for generation and incubation period in the infectiousness factor.
    \item Understand the risk associated with not accounting for users who are asymptomatic.
    \item Compare appropriate decay models for the risk level after a recipient is notified but does not show symptoms (as in equation \ref{eqn:rho-symptom}).
\end{itemize}

Additionally, there is a need to reformulate the risk scoring process in probabilistic terms. While Section \ref{sec:prob-interpretation} does give a probabilistic interpretation of the current risk scoring algorithm, it is preferable to start with an underlying, well-motivated, fully-probabilistic model with clear assumptions. Competing probabilistic models can be consistently evaluated as data is collected.

\printbibliography

\appendix

\section{Visualisations for a single contact event}\label{sec:visualisations-risk}

The risk score for a single contact event factorises as a product of five terms as defined in equation \ref{eqn:indiv-risk-score}. In this section, for visualisation purposes, the risk context adjusting factor $c_{i,m}$ is assumed to be $1$.

In the figures below, we plot visualisations of $r(E_m^i)$ (\verb|risk_score|) as we vary:
\begin{itemize}
    \item the distance $d_{i,m}$ (\verb|distance|),
    \item the time interval $[t_{i,m}]_{days}-[t_i^s]_{days}$ (\verb|time_from_onset|),
    \item the contact event duration $\delta t_{i,m}$ (\verb|duration|).
\end{itemize}

\begin{figure}
    \centering
    \includegraphics[width=1\textwidth]{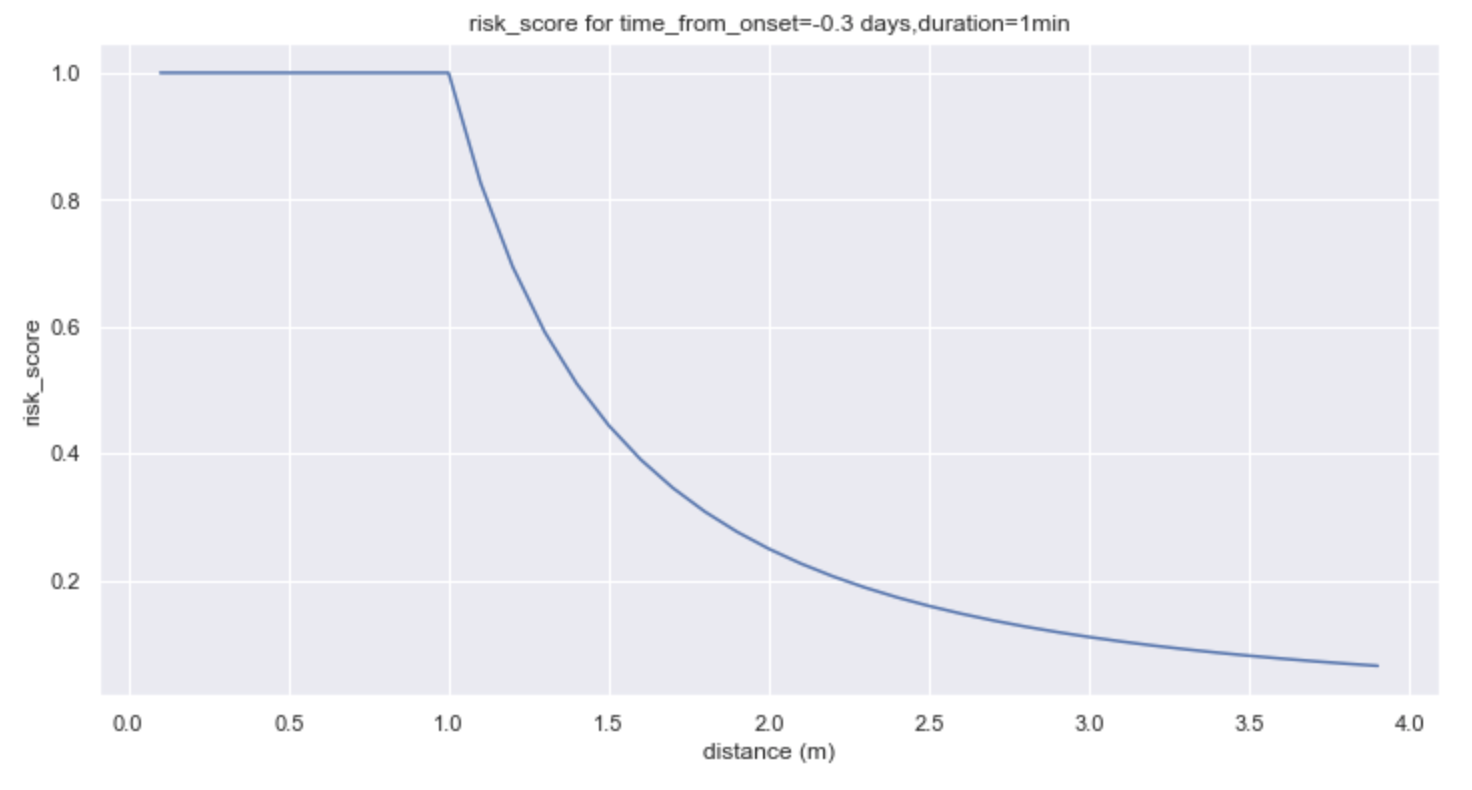}
    \caption{Distance attenuation for constant duration and time from onset}
    \label{fig:risk-distance}
\end{figure}

\begin{figure}
    \centering
    \includegraphics[width=1\textwidth]{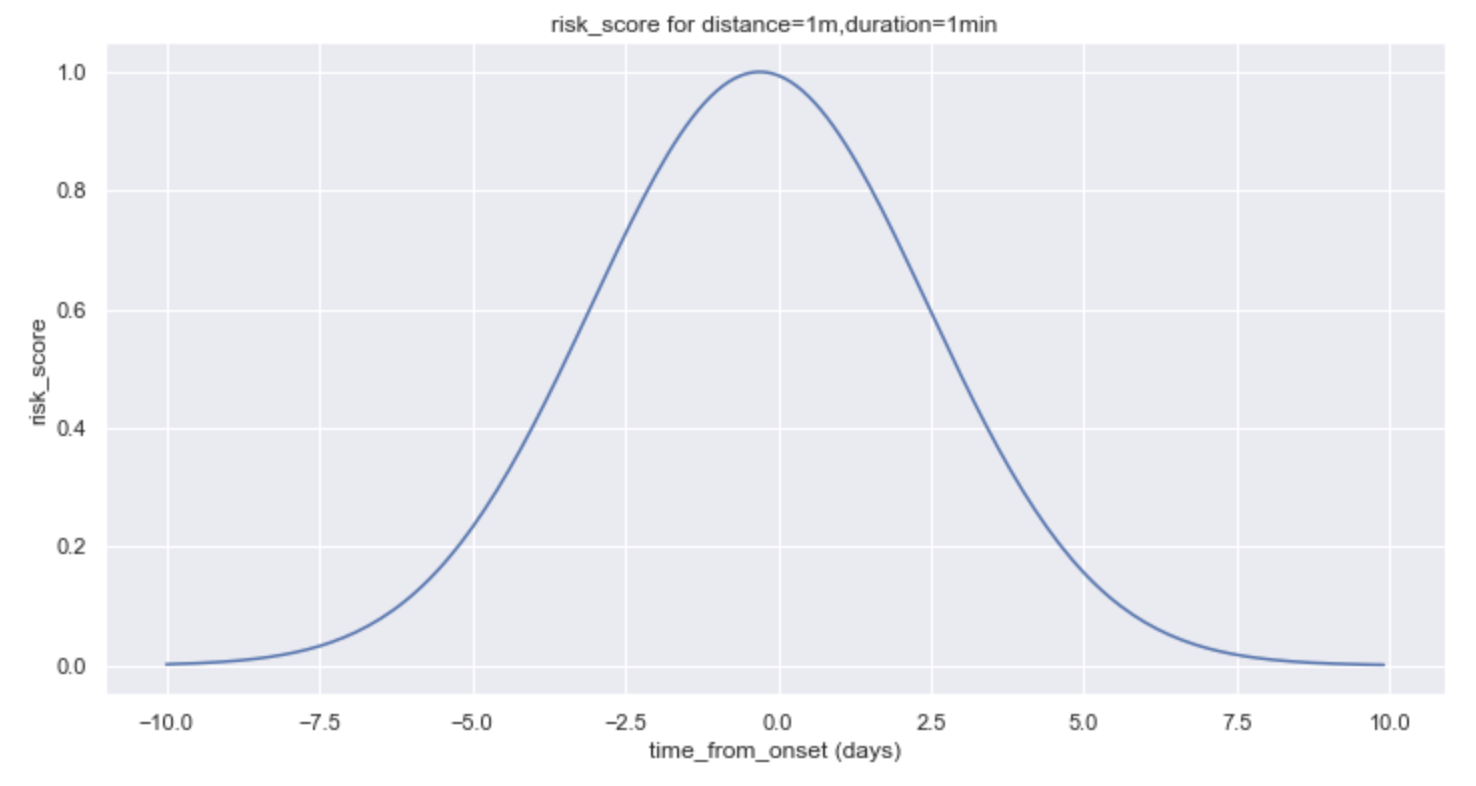}
    \caption{Infectiousness scaling for constant duration and distance}
    \label{fig:risk-infectiousness}
\end{figure}

\begin{figure}
    \centering
    \includegraphics[width=1\textwidth]{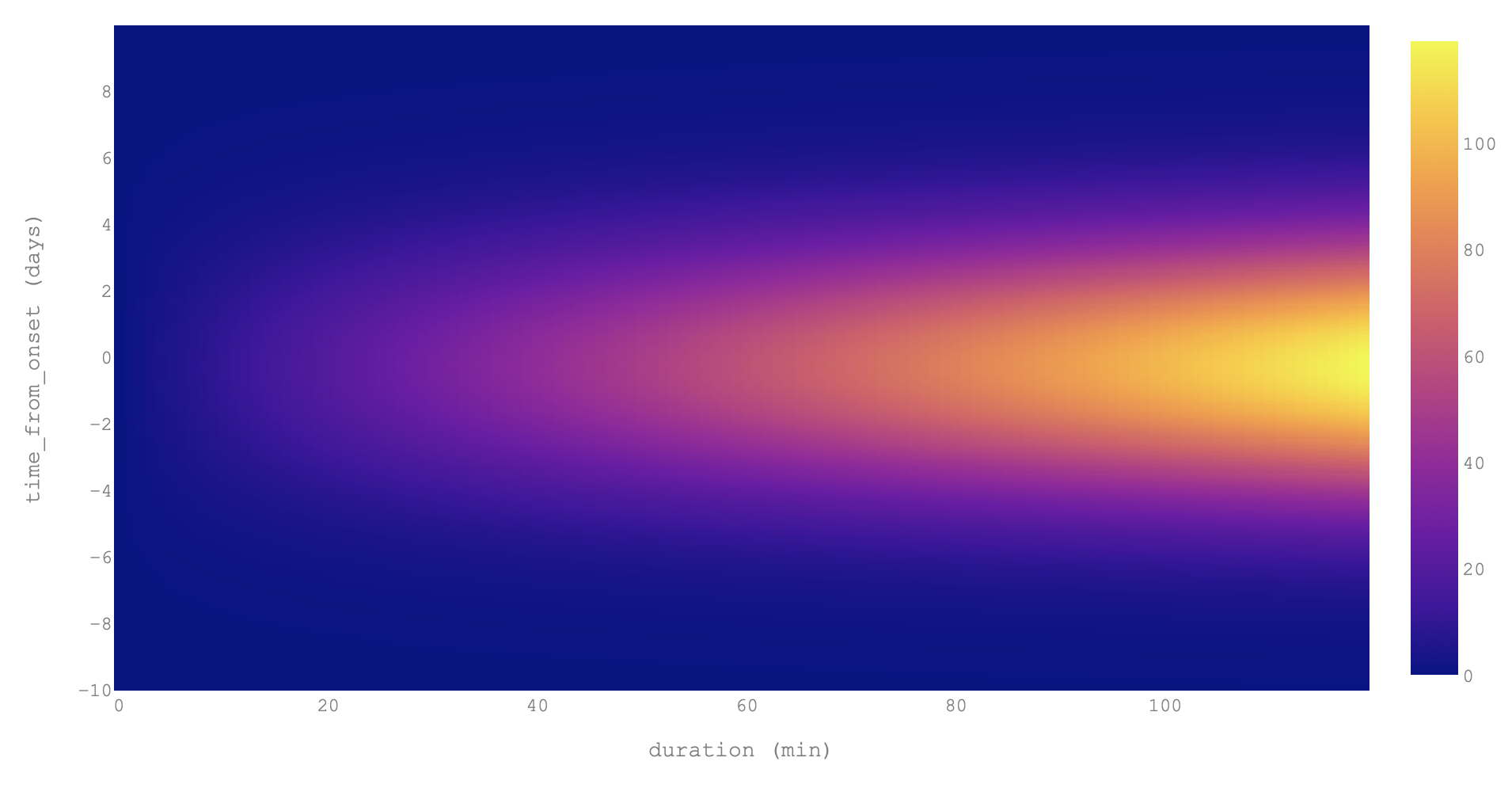}
    \caption{Keeping distance constant at 1m}
    \label{fig:risk-distance-const}
\end{figure}

\begin{figure}
    \centering
    \includegraphics[width=1\textwidth]{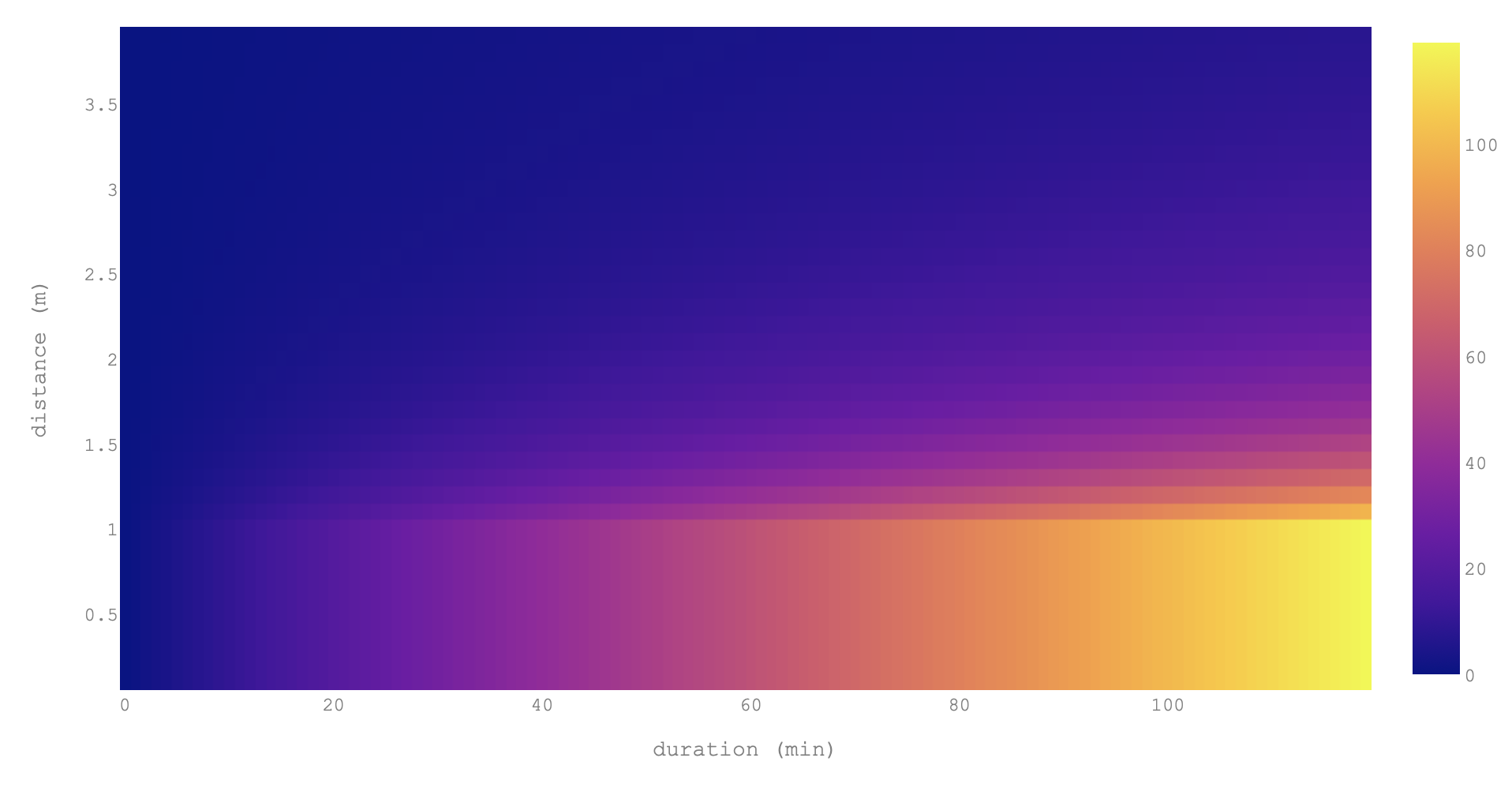}
    \caption{Keeping time from onset constant at -0.3 days}
    \label{fig:risk-onset-const}
\end{figure}

\begin{figure}
    \centering
    \includegraphics[width=1\textwidth]{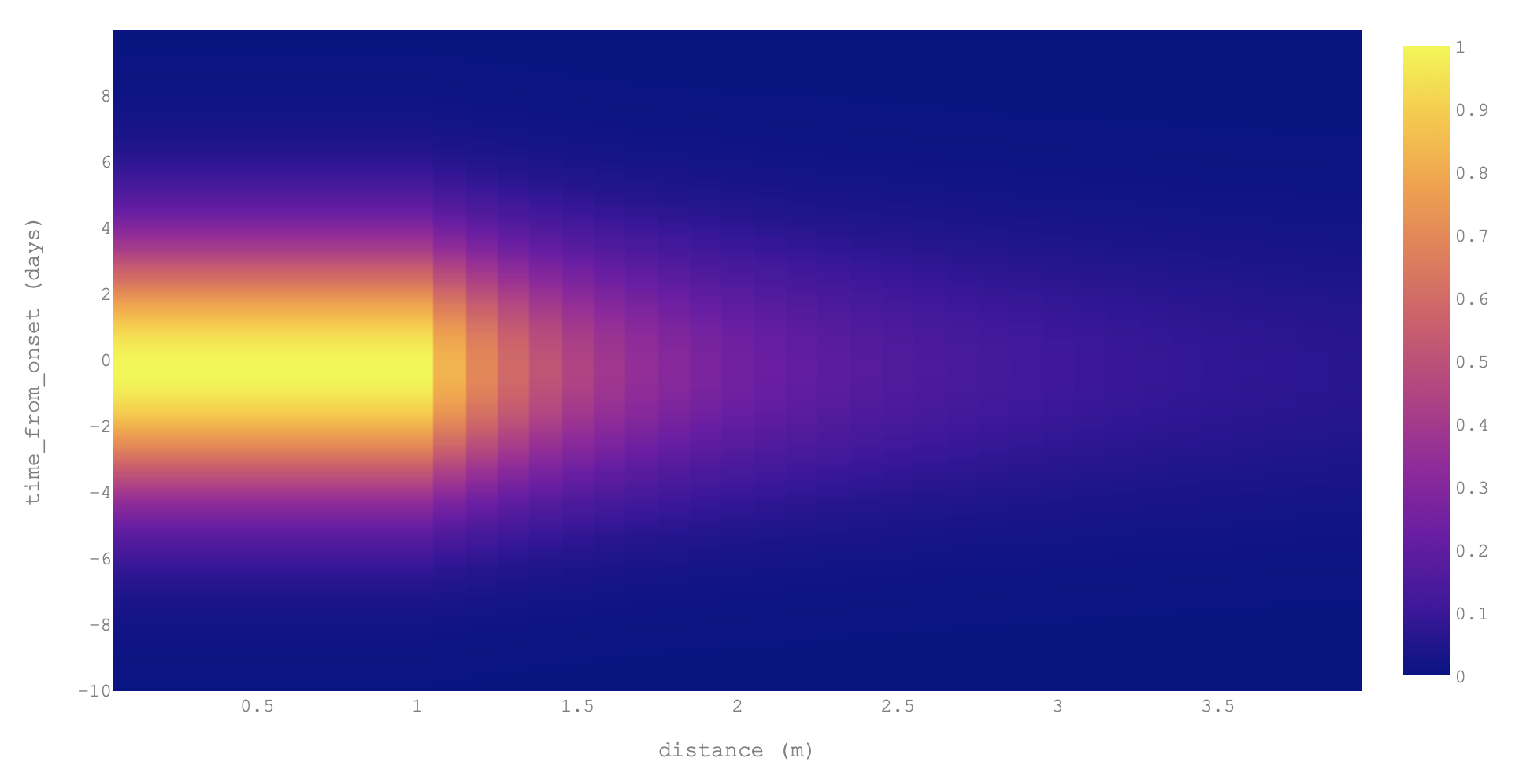}
    \caption{Keeping duration constant at 1 min}
    \label{fig:risk-duration-const}
\end{figure}

\begin{figure}
    \centering
    \includegraphics[width=0.8\textwidth]{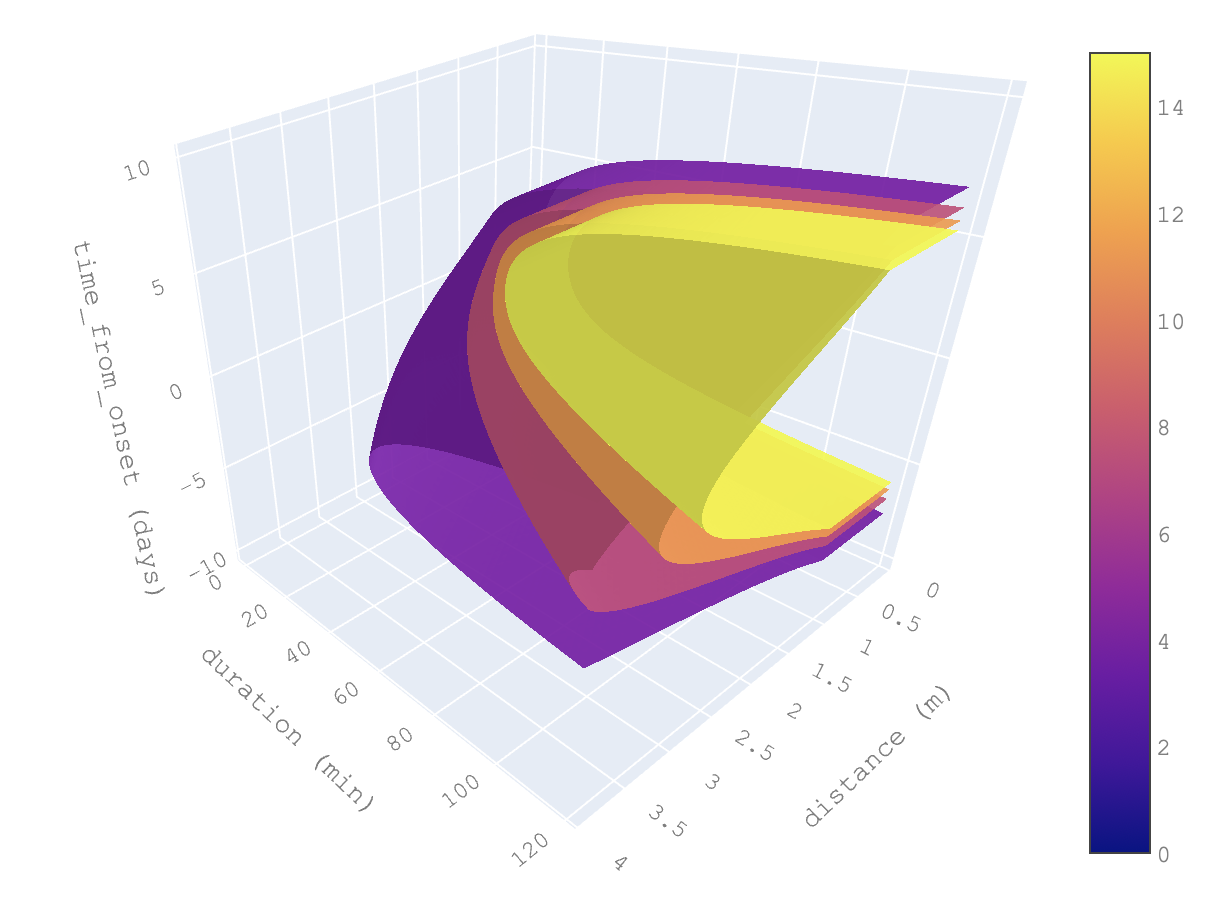}
    \includegraphics[width=0.8\textwidth]{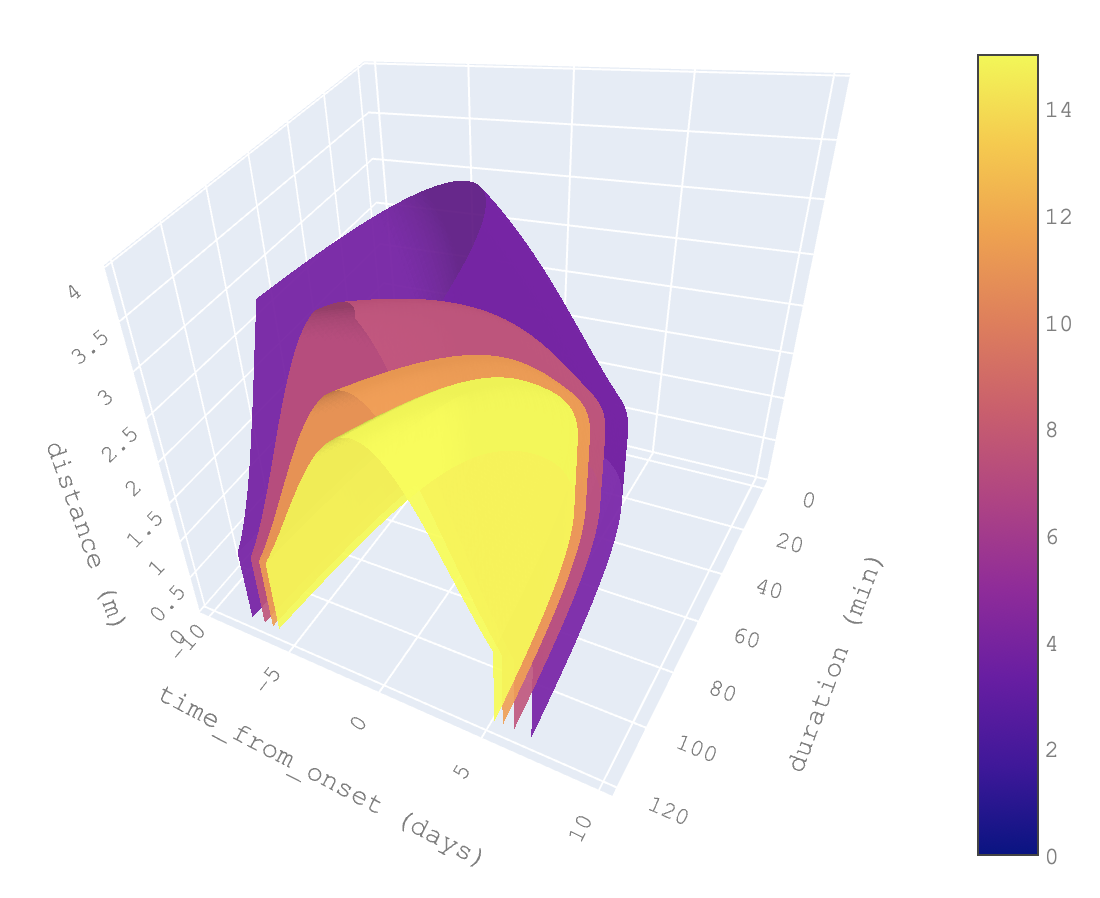}
    \caption{Risk score isosurfaces}
    \label{fig:risk-isosurface}
\end{figure}

Figures \ref{fig:risk-distance}, \ref{fig:risk-infectiousness} show how \verb|risk_score| varies as \verb|distance| and \verb|time_from_onset| vary while keeping the other variables constant. Figures \ref{fig:risk-distance-const}, \ref{fig:risk-onset-const}, \ref{fig:risk-duration-const} show how \verb|risk_score| varies when each of the variables is held constant. Figure \ref{fig:risk-isosurface} plots isosurfaces for \verb|risk_score|.

\section{Derivation of equation \ref{eqn:rho-symptom}}\label{sec:symptom-derivation}

We decompose the probability $\mathbbm{P}(I_{j} | \overline{S(t)})$,
\begin{equation}
\begin{split}
    \mathbbm{P}(I_{j} | \overline{S(t)}) 
    & = \frac{\mathbbm{P}(I_{j}\cap\overline{S(t)})}{\mathbbm{P}(\overline{S(t)})}\\
    & = \frac{\mathbbm{P}(\bigcup_n I_{E_n}\cap\overline{S(t)})}{1-\sum_n\mathbbm{P}(S(t)|I_{E_n})\mathbbm{P}(I_{E_n})}\\
    & = \frac{1 - \prod_n\left(1 - \mathbbm{P}(I_{E_n}\cap\overline{S(t)})\right)}{1-\sum_n\mathbbm{P}(S(t)|I_{E_n})\mathbbm{P}(I_{E_n})} \\
    & = \frac{1 - \prod_n\left(1 - \mathbbm{P}(S(t) | I_{E_n})\mathbbm{P}(I_{E_n})\right)}{1-\sum_n\mathbbm{P}(S(t)|I_{E_n})\mathbbm{P}(I_{E_n})} \\
    & = \frac{1 - \prod_{n}\left(1 - \left( 1 - G(t-t_{E_n})\right)\left(1 - \nu^{\rho(E_n)}\right)\right)}{1-\sum_n G(t-t_{E_n})(1-\nu^{\rho(E_n)})}.
\end{split}
\end{equation}

\end{document}